\documentclass[preprint,showpacs,preprintnumbers,amsmath,amssymb]{revtex4}

\usepackage{amsmath}

\begin{document}



\title{Dynamical symmetries of the Klein-Gordon equation}

\author{Fu-Lin Zhang}
\email[Email:]{flzhang@mail.nankai.edu.cn} \affiliation{Theoretical
Physics Division, Chern Institute of Mathematics, Nankai University,
Tianjin 300071, People's Republic of China}

\author{Jing-Ling Chen}
\email[Email:]{chenjl@nankai.edu.cn}

\affiliation{Theoretical Physics Division, Chern Institute of
Mathematics, Nankai University, Tianjin 300071, People's Republic of
China}

\date{\today}

\begin{abstract}
The dynamical symmetries of the two-dimensional Klein-Gordon
equations with equal scalar and vector potentials (ESVP) are
studied. The dynamical symmetries are considered in the plane and
the sphere respectively. The generators of the $SO(3)$ group
corresponding to the Coulomb potential, and the $SU(2)$ group
corresponding to the harmonic oscillator potential are derived.
Moreover, the generators in the sphere construct the Higgs algebra.
With the help of the Casimir operators, the energy levels of the
Klein-Gordon systems are yielded naturally.
\end{abstract}


\maketitle

\section{Introduction\label{intro}}

Recently, many works about the Dirac or the Klein-Gordon (KG)
equation with  scalar and vector potentials of equal magnitude
(SVPEM) are reported
\cite{ReSymm,Coulomb,Origin,Antinucleon,Hidden,Scarf,Eckart,Pseudospin,equivalent,equivalent2,KGRM,KGCoulomb}
.When the potentials are spherical, the Dirac equation is said to
have the spin or pseudospin symmetry corresponding to the same or
opposite sign. These symmetries, which have been observed in the
hadron and nuclear spectroscopies for a long time
\cite{observe,observe2}, are derived from the investigation of the
dynamics between a quark and an antiquark \cite{quark,generator}.
The very lately studies \cite{equivalent,equivalent2} have revealed
that, the motion of a spin-1/2 particle with SVPEM satisfies the
same differential equation and has the same energy spectrum as a
scalar particle.
When both scalar and vector potentials are spherical, Alberto
\textit{et. al.} \cite{equivalent} have indicated that the
spin-orbit and Darwin terms of either the upper component or the
lower component of the Dirac spinor vanish, which made it
equivalent, as far as energy is concerned, to a spin-0 state. In
this case, besides energy, a scalar particle will also have the same
orbital angular momentum as the (conserved) orbital angular momentum
of either the upper or lower component of the corresponding spin-1/2
particle. These results suggest that, one can image the spin-1/2
particle with SVPEM as a relativistic scalar particle with an
additional spin but without the spin-orbit coupling. From this point
of view, we speculate that, the kinetic characteristics of the Dirac
equation with SVPEM should exist in the KG equation with the same
potentials.

Dynamical symmetries are essential and prevalent both in
non-relativistic classical and quantum mechanics \cite{Greiner}.
Until the work by Ginocchio \cite{U31,U32}, there were no models in
the relativistic quantum mechanics with dynamical symmetry reported.
Ginocchio has found the $U(3)$ and pseudo-$U(3)$ symmetry in the
Dirac equation with SVPEM when the potential takes the harmonic
oscillator form. And in Ref. \cite{Our} we have established the
dynamical symmetries in the two-dimensional Dirac equation (for
hydrogen atom as well as harmonic oscillator) with SVPEM when the
signs are the same (or say ESVP). The goal of this work is to show
the dynamical symmetries of the KG equation with ESVP. Of course,
the following discussion also holds when the signs are opposite, by
making some corrections.

Since there is no an explicit defined Hamiltonian for a spin-$0$
particle, instead we introduce a quasi-Hamiltonian of the KG
equation with ESVP in section \ref{plane}, and consequently provide
an operable criterion for what kind of the dynamical symmetry
exists. The KG equation in a plane with the Coulomb and the harmonic
oscillator form potentials will be discussed as the examples. In
section \ref{sphere}, as a generalization of the plane, we will
investigate the motion of a scalar particle in a sphere with equal
scalar and vector Coulomb or harmonic oscillator potentials.
Conclusion and discussion will be made in the last section.

\section{Dynamical symmetries in a plane \label{plane}}

The KG equation with scalar potential $V_s$ and vector potential
$V_v$ is given by
\begin{eqnarray}\label{KGeq}
\biggr\{p^2+[m+V_s]^2-\biggr[i\frac{\partial}{\partial
t}-V_v\biggr]^2\biggr\}\psi=0.
\end{eqnarray}
For the time-independent potentials, and $V_s= V_v= V(r)/2$, the KG
equation (\ref{KGeq}) becomes
\begin{eqnarray}\label{SV}
[p^2+ (m + \epsilon)V(r)-(\epsilon^2-m^2) ]\psi=0,
\end{eqnarray}
where $\epsilon$ is the relativistic energy.
If we set
\begin{eqnarray}\label{ME}
 \epsilon +m =2\widetilde{m}, \ \ \ \ \epsilon-m=E,
\end{eqnarray}
when $\epsilon \neq -m$, we obtain from Eq. (\ref{SV}) that
\begin{eqnarray}\label{Seq}
\biggr[\frac{p^2}{2 \widetilde{m}} + V(r)- E \biggr]\psi=0.
\end{eqnarray}
When $m\rightarrow \infty$, $\epsilon \rightarrow m$, Eq.
(\ref{Seq}) returns to the usual Schr\"{o}dinger equation.

Suppose $U$ is a operator of a symmetry group of the spin-$0$ system
with ESVP, with the generators denoted as $L_i$. Generally, if
$\psi$ fulfills Eq. (\ref{Seq}), the state $\psi'=U \psi$ should
also fulfills it, namely
\begin{eqnarray}\label{Sequ}
\biggr[\frac{p^2}{2 \widetilde{m}} + V(r)- E \biggr] U \psi=0.
\end{eqnarray}
Multiplying Eq. (\ref{Seq}) by $U$ from the left-hand side and
comparing it with Eq. (\ref{Sequ}), we obtain
\begin{eqnarray}\label{com}
[U,\widetilde{H}]=U\widetilde{H}-\widetilde{H}U=0,
\end{eqnarray}
where $\widetilde{H}=\frac{p^2}{2 \widetilde{m}} + V(r)$ is called
the \textit{quasi-Hamiltonian} in this paper. Immediately, the
generators of the symmetry group $L_i$ also commute with
$\widetilde{H}$
\begin{eqnarray}\label{comL}
[L_i,\widetilde{H}]=0.
\end{eqnarray}
This result provide an operable criterion for what kind of the
dynamical symmetry exists in the KG equation with ESVP.

As examples, let us consider the motion of a scalar particle
constrained in a two-dimensional (2D) plane.

\textit{Coulomb potential}. When the potential takes the Coulomb
form $V(r)=-\frac{k}{r}$, then the quasi-Hamiltonian reads
\begin{eqnarray}\label{Hhydrongen}
\widetilde{H}=\frac{p^2}{2\widetilde{m}}-\frac{k}{r}.
\end{eqnarray}
It is easy to obtain the generators from the non-relativistic
results as
\begin{eqnarray}\label{Ghydrongen}
L &=& x_1 p_2 -x_2 p_1, \nonumber \\
\widetilde{R}_1 &=& \frac{1}{2\widetilde{m}k}(Lp_2+p_2L)-\frac{x_1}{r},\\
\widetilde{R}_2 &=&
\frac{1}{2\widetilde{m}k}(-Lp_1-p_1L)-\frac{x_2}{r},\nonumber
\end{eqnarray}
which commute with $\widetilde{H}$, and satisfy the commutation
relations
\begin{eqnarray}\label{Chydrongen}
\lbrack L,\widetilde{R}_1 \rbrack=i\widetilde{R}_2,\ \lbrack
L,\widetilde{R}_2 \rbrack=-i\widetilde{R}_1,\ \lbrack
\widetilde{R}_1,\widetilde{R}_2
\rbrack=\frac{-i2\widetilde{H}}{\widetilde{m}k^2}L,
\end{eqnarray}
and $\widetilde{R}_1 ^2+\widetilde{R}_2
^2=2\frac{\widetilde{H}}{\widetilde{m}k^2}(L^2+\frac{1}{4})+1$.
These results show that the 2D KG equation with equal scalar and
vector potentials has the $SO(3)$ symmetry based on the above
criterion. The relations of the generators can be also used to
determine the energy levels of this system. After defining the
normalized generators
\begin{eqnarray}\label{A}
A_3 = L,\ \ A_i =
\biggr[-\frac{2E}{\widetilde{m}k^2}\biggr]^{-\frac{1}{2}}\widetilde{R}_i,\;\;
(i=1,2),
\end{eqnarray}
one then obtains
\begin{eqnarray}\label{ACR}
[A_i,A_j]=i\epsilon_{ijk}A_k, \;\;(i,j,k=1,2,3).
\end{eqnarray}
The $SO(3)$ Casimir operator is given by
\begin{eqnarray}\label{CSO3}
C_{so3}=A^2_1+A^2_2+A^2_3=j(j+1),\ \ \ j=0,1,2,...
\end{eqnarray}
Inserting Eq. (\ref{A}) into Eq. (\ref{CSO3}), one can have
\begin{eqnarray}\label{Ehn}
E=-\frac{2k^2}{n^2}\widetilde{m},\ \ \ n=2j+1=1,3,5,...
\end{eqnarray}
Considering the relation in Eq. (\ref{ME}), the relativistic energy
levels of this system are given by
\begin{eqnarray}\label{Eh}
\epsilon=E+m=\frac{n^2-k^2}{n^2+k^2}m,
\end{eqnarray}
which coincides with the results in 2D Dirac system \cite{Our}.

In the non-relativistic limit ($m \rightarrow \infty$, $\epsilon
\rightarrow m$),  $\widetilde{m}$ and $\widetilde{H}$ reduce to the
mass $m$ and the Hamiltonian of the non-relativistic hydrogen atom,
respectively. In the meantime, the generators $\widetilde{R}_i$
reduce to the components of the Rung-Lenz vector, and Eq.
(\ref{Ehn}) is nothing but the spectrum of a non-relativistic
hydrogen atom.

\textit{Harmonic oscillator potential}. When the potential takes the
harmonic oscillator form, $V(r)=\frac{1}{2}m\omega^2r^2$, in
comparison with the non-relativistic harmonic oscillator, we set
\begin{eqnarray}\label{omg}
\frac{1}{2}\widetilde{m}\widetilde{\omega}^2=\frac{1}{2}m \omega^2,\
\ \ \widetilde{\omega}=\sqrt{\frac{m \omega^2}{\widetilde{m}}},
\end{eqnarray}
then the quasi-Hamiltonian becomes
\begin{eqnarray}\label{Ho}
\widetilde{H}=\frac{p^2}{2\widetilde{m}} +
\frac{1}{2}\widetilde{m}\widetilde{\omega}^2 r^2.
\end{eqnarray}
For a fixed energy eigenvalue, $\widetilde{m}$ and
$\widetilde{\omega}$ are constants. Thus the generators commuting
with $\widetilde{H}$ are given by
\begin{eqnarray}\label{Go}
J_1 &=& \frac{1}{2}\biggr(\frac{1}{\widetilde{m}\widetilde{\omega}}p_1p_2+\widetilde{m}\widetilde{\omega}x_1x_2\biggr),\nonumber\\
J_2 &=& \frac{1}{2}\biggr(x_1p_2-x_2p_1\biggr),\\
J_3 &=&
\frac{1}{2}\biggr(\frac{1}{\widetilde{m}\widetilde{\omega}}\frac{p_1^2-p_2^2}{2}
+\widetilde{m}\widetilde{\omega}\frac{x_1^2-x_2^2}{2}\biggr).\nonumber
\end{eqnarray}
They satisfy the $SU(2)$ commutation relations
\begin{eqnarray}\label{JCR}
[J_i,J_j]=i\epsilon_{ijk}J_k, (i,j,k=1,2,3).
\end{eqnarray}
The Casimir operator of the $SU(2)$ group is
\begin{eqnarray}\label{Csu2}
C_{su2} &=& J_1^2+J_2^2+J_3^2=
\frac{1}{4}\biggr(\biggr(\frac{\widetilde{H}}{\widetilde{\omega}}\biggr)^2-1\biggr) \nonumber\\
&=&s(s+1),\ s=0,\frac{1}{2},1...
\end{eqnarray}
which yields the energy levels as
\begin{eqnarray}\label{Eo}
E=(n+1)\widetilde{\omega},\ \ n=2s=0,1,2...
\end{eqnarray}
Combining with the relations (\ref{omg}) and (\ref{ME}), it is
straightforward to prove the relativistic energy is the real root of
the cubic equation
\begin{eqnarray}
(\epsilon-m)^2(\epsilon+m)=2m\omega^2(n+1)^2,
\end{eqnarray}
which is the same as the spectrum of the Dirac equation given in
\cite{Our}. In non-relativistic limit, $\widetilde{m}\rightarrow m$
and $\widetilde{\omega}\rightarrow \omega$, the above formulae
return to the results of the non-relativistic harmonic oscillator.

\section{Dynamical symmetries in a sphere \label{sphere}}
In recent years, polynomial angular momentum algebra and its
increasing applications have been the focus of very active research.
The first special case, called the Higgs algebra now, is found by
Higgs \cite{Higgs} in the motion of a non-relativistic particle in a
2D curved space with the Coulomb or harmonic oscillator potential.
It have been shown in the above section that, the KG system with
ESVP has the same dynamical symmetry as the non-relativistic system
with the similar potential. This gives rise to an interesting
question: Does the Higgs algebra exists in the KG equation in a
sphere with ESVP, when the potential takes the Coulomb or harmonic
oscillator form?

To solve this problem, we first construct the classical Hamiltonian
for a relativistic paticle in a sphere. In the gnomonic projection,
as introduced in \cite{Higgs}, the metric is given by
\begin{eqnarray}\label{metric}
(ds)^2=\frac{d\vec{x} \cdot d\vec{x}}{1+\lambda r^2} - \frac{\lambda
(\vec{x}\cdot d\vec{x})^2}{(1+\lambda r^2)^{2}},
\end{eqnarray}
where $\lambda$ is the curvature of the sphere. The Lagrangian for
the free motion of a relativistic scalar particle is
$\mathcal{L}=-m\sqrt{1-\dot{s}^2}$, where $m$ is the mass and
$\dot{s}^2$ is defined in Eq. (\ref{metric}). The momentum conjugate
to $x_i$ ($i=1,2$) is
\begin{eqnarray}\label{p}
p_i=-\frac{m^2}{\mathcal{L}}\biggr[\frac{\dot{x}_i}{1+\lambda r^2}-
\frac{\lambda (\vec{x}\cdot \dot{\vec{x}})x_i}{(1+\lambda
r^2)^{2}}\biggr],
\end{eqnarray}
and the Hamiltonian is given by
\begin{eqnarray}\label{HKG}
H^2=m^2+\pi^2+\lambda L^2,
\end{eqnarray}
where $L=x_1p_2-x_2p_1$, $\pi^2=\pi_1^2+\pi_2^2$, and
\begin{eqnarray}\label{pi}
\pi_i=p_i+\frac{1}{2}\lambda[x_i(\vec{x}\cdot\vec{p})+(\vec{p}\cdot\vec{x})x_i],\
\ \ i=1,2.
\end{eqnarray}
Thus, in the relativistic quantum mechanics, the KG equation of a
free particle in a sphere is
\begin{eqnarray}\label{KGfree}
-\frac{\partial^2}{\partial t^2} \psi = (m^2+\pi^2+\lambda L^2)\psi.
\end{eqnarray}

Suppose the particle coupling to a scalar potential $V_s$ and a
vector potential $V_v$ (only the time component is nonzero), and
$V_s=V_v=\frac{V(r)}{2}$, the KG equation (\ref{KGfree}) becomes
\begin{eqnarray}\label{KGv}
\biggr[\frac{1}{2 \widetilde{m}}(\pi^2+\lambda L^2) + V(r)- E
\biggr]\psi=0,
\end{eqnarray}
on the premise that the energy $\epsilon \neq -m$, and
$\widetilde{m}$ and $E$ take the definitions in Eq. (\ref{ME}).
Then, the quasi-Hamiltonian in a sphere is defined as
$\widetilde{H}=\frac{1}{2 \widetilde{m}}(\pi^2+\lambda L^2) + V(r)$.
Unlike the equations (\ref{SV}) and (\ref{Seq}), the KG equation in
a sphere with ESVP hasn't an equivalent Dirac equation, because the
spin-orbit coupling is kept by the curvature.

\textit{Coulomb potential}. When $V(r)=-\frac{k}{r}$, the
quasi-Hamiltonian $\widetilde{H}=\frac{1}{2
\widetilde{m}}(\pi^2+\lambda L^2) -\frac{k}{r}$ commutes with the
angular momentum $L$ and
\begin{eqnarray}
\widetilde{R}_1 &=& \frac{1}{2\widetilde{m}k}(L\pi_2+\pi_2L)-\frac{x_1}{r},\\
\widetilde{R}_2 &=&
\frac{1}{2\widetilde{m}k}(-L\pi_1-\pi_1L)-\frac{x_2}{r}.\nonumber
\end{eqnarray}
Set $\widetilde{R}_{\pm}=\widetilde{R}_1 \pm i\widetilde{R}_2$, one
can find the commutators
\begin{eqnarray}
\lbrack L,\widetilde{R}_{\pm} \rbrack = \pm
\widetilde{R}_{\pm},\;\;\; \lbrack
\widetilde{R}_{+},\widetilde{R}_{-}\rbrack = c_3L^3+c_1L,
\end{eqnarray}
where $c_3=\frac{4\lambda}{\widetilde{m}^2 k^2}$ and
$c_1=-\frac{4\widetilde{H}}{\widetilde{m}
k^2}+\frac{\lambda}{2\widetilde{m}^2 k^2}$. These relations indicate
that the generators construct the Higgs algebra, and the system has
the $SO(3)$ symmetry. To get the energy levels, we should calculate
the anticommutator
\begin{eqnarray}
\{ \widetilde{R}_{+},\widetilde{R}_{-} \} =
2+\frac{\widetilde{H}}{\widetilde{m}
k^2}
+\frac{(8\widetilde{m}\widetilde{H}-5 \lambda)L^2}{2\widetilde{m}^2
k^2} -\frac{2 \lambda L^4}{\widetilde{m}^2 k^2}.
\end{eqnarray}
The Casimir of the Higgs algebra is given by \cite{CofH,XJZ}
\begin{eqnarray}
C&=& \{ \widetilde{R}_{+},\widetilde{R}_{-}
\}+\biggr(c_1+\frac{c_3}{2}\biggr)L^2+\frac{c_3}{2}L^4 \nonumber \\
&=& 2+\frac{\widetilde{H}}{\widetilde{m} k^2}\nonumber \\
&=& c_1\; C_{so3}+\frac{c_3}{2}\; C_{so3}^2,
\end{eqnarray}
where $C_{so3}=j(j+1)$, $j=0,1,2,...$, is the $SO(3)$ Casimir
operator. Based on which, the energy eigenvalues satisfy
\begin{eqnarray}\label{Ehs}
\epsilon-m=-\frac{\epsilon+m}{(2j+1)^2}k^2+\frac{\lambda}{\epsilon+m}j(j+1).
\end{eqnarray}
When $\lambda \rightarrow 0$, it reduces to the result in the plane
as Eq. (\ref{Eh}). In the non-relativistic limit, when $m
\rightarrow \infty$ and $\epsilon \rightarrow m$, Eq. (\ref{Ehs})
leads to the energy levels given in \cite{Higgs}.

\textit{Harmonic oscillator potential}. For the harmonic oscillator
potential $V(r)=\frac{1}{2}m\omega^2r^2$, we also deal with it by
using Eq. (\ref{omg}). Then, the quasi-Hamiltonian becomes
\begin{eqnarray}
\widetilde{H}=\frac{1}{2 \widetilde{m}}(\pi^2+\lambda L^2)
+\frac{1}{2}\widetilde{m}\widetilde{\omega}^2 r^2,
\end{eqnarray}
which commutes with the angular momentum $L$ and the second order
tensors
\begin{eqnarray}
\widetilde{s}_1 &=& \frac{1}{\widetilde{m}
\widetilde{\omega}}\frac{\pi_1\pi_2+\pi_2\pi_1}{2}+\widetilde{m}
\widetilde{\omega} x_1 x_2,\nonumber \\
\widetilde{s}_2 &=& \frac{1}{\widetilde{m}
\widetilde{\omega}}\frac{\pi_1^2-\pi_2^2}{2}+\widetilde{m}
\widetilde{\omega} \frac{x_1^2-x_2^2}{2}.
\end{eqnarray}
Set $J_{\pm}=\frac{1}{2}(\widetilde{s}_2 \pm i \widetilde{s}_1)$ and
$J_{3}=\frac{1}{2}L$. They satisfy the Higgs algebra relations
$\lbrack J_3,J_{\pm} \rbrack = \pm J_{\pm}$, $\lbrack J_+,J_{-}
\rbrack = a_3 J_3^3+ a_1 J_3$, with
$a_1=2(1-\frac{1}{4}\frac{\lambda^2}{\widetilde{m}^2\widetilde{\omega}^2}
+\frac{\lambda}{\widetilde{m}\widetilde{\omega}^2}\widetilde{H})$
and $a_3=-4\frac{\lambda^2}{\widetilde{m}^2\widetilde{\omega}^2}$.

The anticommutation relation is
\begin{eqnarray}
\{
J_+,J_-\}&=&2\frac{\lambda^2}{\widetilde{m}^2\widetilde{\omega}^2}J_3^4
+\biggr(-2\frac{\lambda}{\widetilde{m}}\widetilde{H}
-2+\frac{5}{2}\frac{\lambda^2}{\widetilde{m}^2}\widetilde{\omega}^2\biggr)J_3^2
 +\biggr(\frac{\widetilde{H}^2}{2\widetilde{\omega}^2}-\frac{1}{2}
-\frac{\lambda}{2\widetilde{m}\widetilde{\omega}^2}\widetilde{H}\biggr),
\end{eqnarray}
and the Casimir of the Higgs algebra reads
\begin{eqnarray}
C &=& \{ J_+,J_- \}+\biggr(a_1+\frac{a_3}{2}\biggr)J_3^2+\frac{a_3}{2}J_3^4 \nonumber \\
  &=&
  \frac{\widetilde{H}^2}{2\widetilde{\omega}^2}-\frac{1}{2}-\frac{\lambda}{2\widetilde{m}\widetilde{\omega}^2}\widetilde{H}\nonumber \\
  &=& a_1C_{su2}+\frac{a_3}{2}C_{su2}^2,
\end{eqnarray}
where $C_{su2}=s(s+1)$, $s=0,\frac{1}{2},1...$, is the $SU(2)$
Casimir operator. Thus, we obtain the equation which the energy
levels satisfy as
\begin{eqnarray}\label{Eos}
\epsilon- m =\frac{\lambda
(n+1)^2}{\epsilon+m}+\sqrt{\frac{2m}{\epsilon+m}\omega^2+\frac{\lambda^2}{(\epsilon+m)^2}}(n+1),
\end{eqnarray}
where $n=2s=0,1,2,...$. This result showes that the KG equation with
equal scalar and vector harmonic oscillator potential has the
$SU(2)$ symmetry. When $\lambda \rightarrow 0$, Eq. (\ref{Eos})
reduces to the energy in Eq. (\ref{Eo}). And the non-relativistic
limit give the result in \cite{Higgs}, when $m \rightarrow \infty$
and and the coefficient of elasticity $m\omega^2$ keeps
unchangeably.

\section{Conclusion and discussion\label{conclu}}
To discuss the symmetry of the K-G equation with ESVP, we have
introduced a quasi-Hamiltonian $\widetilde{H}$. The generators of
the symmetry group commute with $\widetilde{H}$. We have
investigated the motion of a relativistic scalar particle both in a
plane and a sphere by providing some examples. The symmetry is shown
to be the $SO(3)$ for the Coulomb potential and the $SU(2)$ for the
harmonic oscillator potential. Specially, in the sphere, the
generators according to the two types of potentials construct the
Higgs algebra respectively. The Casimir operators of these systems
can be used to calculate the energy spectra straightway.

The procedure, to study the dynamical symmetry of the KG with ESVP
in this work, is applicable in not only 2D systems but also
three-dimensional or N-dimensional (ND) systems. We can foretell the
ND KG equation with ESVP has the $SO(N+1)$ symmetry for the Coulomb
potential and the $SU(N)$ for harmonic potential. These results
exhibit a new visual angle to understand the dynamical symmetries in
the Dirac systems with spin or pseudospin symmetry. From the results
of the above examples, we can put forward a uniform approach to
solve the spectra of the KG equation or the Dirac equation with
ESVP, i.e., if the Schr\"{o}dinger equation with a certain potential
is integrable, the energy level is expressed as a function of the
mass $m$, the parameters $\{k_{\alpha}\}$ involving in the potential
$V(r)$ and a set of good quantum numbers $\{n_j\}$ as
$\varepsilon=\varepsilon(m,k_{\alpha},n_j)$, then the corresponding
relativistic energy spectra of the KG equation with ESVP satisfies
$\epsilon-m=\varepsilon(\frac{1}{2}(\epsilon+m),k_{\alpha},n_j)$.

\begin{acknowledgments}
We thank W.-S. Dai and M. Xie for their valuable discussions. This
work is supported in part by NSF of China (Grants No. 10575053 and
No. 10605013), Program for New Century Excellent Talents in
University, and the Project-sponsored by SRF for ROCS, SEM.
\end{acknowledgments}

\bibliography{KGeq-1}

\end{document}